\documentclass[11pt]{article}
\usepackage{graphicx}
\usepackage{epsfig}

\RequirePackage{xspace}

\def\Bbar    {\kern 0.18em\overline{\kern -0.18em B}{}\xspace}
\def\Dbar    {\kern 0.18em\overline{\kern -0.18em D}{}\xspace}
\def\ubar    {\kern 0.18em\overline{\kern -0.18em u}{}\xspace}
\def\cbar    {\kern 0.18em\overline{\kern -0.18em c}{}\xspace}

\def\BBbar{\mbox{$B\overline {B}\ $}}
\def\Bz      {\ensuremath{B^0}\xspace}

\def\Bm      {\ensuremath{B^-}\xspace}
\def\Bzb     {\ensuremath{\Bbar^0}\xspace}
\def\K0S         {\ensuremath{K^0_S}\xspace}
\def\CP                {\ensuremath{C\!P}\xspace}
\def\CKM                {\ensuremath{C\!K\!M}\xspace}
\def\UT                {\ensuremath{U\!T}\xspace}
\def\SM                {\ensuremath{S\!M}\xspace}
\def\NP                {\ensuremath{N\!P}\xspace}

\def\Dz      {\ensuremath{D^0}\xspace}
\def\Dp      {\ensuremath{D^+}\xspace}

\newcommand{\mes}{\ensuremath{{m_{ES}}}} 
\newcommand{\DE}{\ensuremath{\Delta E}}
\newcommand{\deltab}{\ensuremath{\delta_B}}
\newcommand{\deltas}{\ensuremath{\delta_s}}
\def \deltabst   {\ensuremath{\delta^*_B}\xspace}

\def\Nubar    {\kern 0.18em\overline{\kern -0.18em \nu}{}\xspace}

\def\ra                 {\ensuremath{\rightarrow}\xspace}

\def\babar{{\em B}{\footnotesize\em A}{\em B}{\footnotesize\em AR}}

\newcommand{\NNbtoappim}{\mbox{$B^0 \rightarrow a^{\pm}_1\, \pi^{\mp}  $}}

\def\Kuno  {\ensuremath{K_1(1270)}}
\def\Kunop {\ensuremath{K_1(1400)}}

\newcommand{\Nbtoappim}{\mbox{$B^0 \rightarrow a_1(1260)^{\pm}\, \pi^{\mp}  $}}
\newcommand{\aunob}{\mbox{$a_1$}}
\def\bbar  {\mbox{$\bar{b} $}}

\def\rhoprhom{\mbox{$\rho^+ \rho^-$}}
\def\rhozrhoz{\mbox{$\rho^0 \rho^0$}}
\def\rhoprhoz{\mbox{$\rho^+ \rho^0$}}

\newcommand{\aunoppim}{\mbox{$a_1^+ \pi^-$}}
\newcommand{\aunompip}{\mbox{$a_1^- \pi^+$}}

\def\ubaru{\mbox{$\bar{u}  u $}}
\def\dbar  {\mbox{$\bar{d} $}}

\newcommand{\etal}{{\em et al.}}
\newcommand{\jprlBase}       {Phys.\ Rev.\ Lett.\xspace}
\newcommand{\jprl}      [1]  {\jprlBase\ {\bf #1}}
\newcommand{\jprBase}        {Phys.\ Rev.\xspace}
\newcommand{\jprd}      [1]  {\jprBase\ D~{\bf #1}}

\newcommand{\jplBase}        {Phys.\ Lett.\xspace}
\newcommand{\plb}       [1]  {\jplBase\ B~{\bf #1}}

\newcommand{\jnpBase}        {Nucl.\ Phys.\xspace}
\newcommand{\npb}       [1]  {\jnpBase\ B~{\bf #1}}
\newcommand{\npa}       [1]  {\jnpBase\ A~{\bf #1}}

\newcommand{\nimBaseA}       {Nucl.\ Instr.\ Meth.\xspace}
\newcommand{\nima}      [1]  {\nimBaseA~A~{\bf #1}}

\newcommand{\epjBase}       {Eur.\ Phys.\ J.\xspace}
\newcommand{\epjc}      [1]  {\epjBase\ C~{\bf #1}}

\newcommand{\jhepBase}       {JHEP\xspace}
\newcommand{\jhep}      [1]  {\jhepBase\ ~{\bf #1}}

\newcommand{\progtpBase}       {Prog.\ Theor.\ Phys.\xspace}
\newcommand{\progtp}      [1]  {\progtpBase\ ~{\bf #1}}

\newcommand{\jpBase}       {J.\ Phys.\xspace}
\newcommand{\jp}      [1]  {\jpBase\ G ~{\bf #1}}

\newcommand{\etapr}{\ensuremath{\eta^\prime}}			
\newcommand{\psiKs}{\mbox{$B^0\ra J/\psi  K^0_S $}}
\newcommand{\gevcc}{\ensuremath{{\mathrm{\,Ge\kern -0.1em V\!/}c^2}}\xspace}
\newcommand{\gevc}{\ensuremath{{\mathrm{\,Ge\kern -0.1em V\!/}c}}\xspace}
\newcommand{\gevccq}{\ensuremath{{\mathrm{\,Ge\kern -0.1em V^2\!/}c^4}}\xspace}

\begin{document}
{\pagestyle{empty}

\begin{flushright}
\end{flushright}

\par\vskip 4cm

\begin{center}
\Large \bf  $B$ Decays and CP Violation  from  \babar
\end{center}
\bigskip

}
\begin{center}
\large 
F. Palombo\\
Universit\`a degli Studi di Milano, Dipartimento di Fisica and INFN, I-20133
Milano, Italy \\
{\it on behalf of the \babar\  Collaboration}
\end{center}
\bigskip \bigskip

\begin{center}
\large \bf Abstract
\end{center}
We present  some recent \babar\  measurements of the magnitudes  of  the elements $V_{ub}$ and $V_{cb}$ of the Cabibbo-Kobayashi-Maskawa quark-mixing matrix,  and of the  angles $\alpha$ and $\gamma $ of the unitary triangle of the standard model of the electroweak interactions.
 Most of the  measurements presented here are based  on the full   \babar\ 
$\Upsilon(4S)$ dataset, consisting of about  $467 \times 10^{6}$ \BBbar  pairs.    

\vfill
\begin{center}
Invited talk  presented  at  the 8$^{th}$ Latin American Symposium on High Energy Physics , \\
6/12/2010---12/12/2010, Valparaiso, Chile
\end{center}

\newpage%

\section{Introduction}
The elements of the Cabibbo-Kobayashi-Maskawa (\CKM) quark-mixing matrix~\cite{CKM} are fundamental parameters
of the Standard Model (\SM) of electroweak interactions.  \CKM\   matrix  is determined by four independent  parameters,
interpreted as three  mixing angles between the three pairs of quark generations and a non-trivial complex phase which in the \SM\ represents the only source  of Charge-Parity (\CP) violation.  

The unitarity condition of  the  \CKM\  matrix leads to six relations, which  represent six triangles in the complex plane. Four of these triangles are degenerate with one side much smaller than the other two and are not useful in the present  experimental sensitivity.  The remaining two triangles have the lengths  of all  sides  of  order $\lambda^3$, where $\lambda$ is the sine of the Cabibbo angle ($\lambda \sim 0.225)$.   To leading order in $\lambda$, these two triangles coincide.  The \CKM\ Unitary Triangle (\UT) is  taken the one that represents the relation   $V^*_{ub} V_{ud} + V^*_{cb} V_{cd} + V^*_{tb} V_{td}  = 0$.   This triangle can be rescaled  \cite{Rescaled} in order to have one side of unitary length on one axis as  shown in Fig.~\ref{fig:triangolo}:

\begin{figure}[!h]
\vspace*{0.5cm}
\hspace*{-0.5cm}
\begin{center}
\includegraphics[angle=0, scale=0.333] {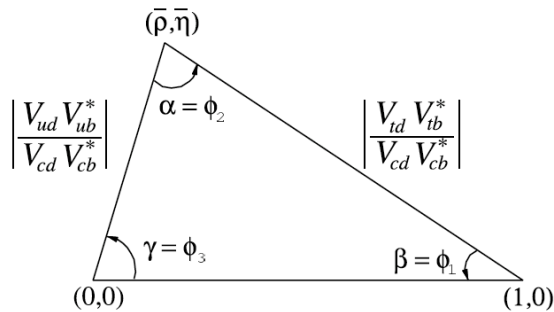} 
\vspace*{-0.5cm}
\caption{ Unitary  Triangle}
\end{center}
\label{fig:triangolo}
\end{figure}

The lengths of the other two sides are :

\begin{equation}
R_u \equiv \left | \frac{V_{ud}V^*_{ub}}{V_{cd}V^*_{cb}} \right | =  (1-\frac{\lambda^2}{2}) \frac{1}{\lambda} \left | \frac{V_{ub}}{V_{cb}} \right |,  \quad    R_t \equiv \left | \frac{V_{td}V^*_{tb}}{V_{cd}V^*_{cb}} \right | = \frac{1}{\lambda} \left | \frac{V_{tb}}{V_{cb}} \right | \\
\label{equ=sides}
\end{equation}
 
 while the  angles \cite{Notazione} are defined  by:
 
 \begin{equation}
\alpha \equiv arg\left [ - \frac{V_{td}V^*_{tb}}{V_{ud}V^*_{ub}} \right ],  \quad    \beta \equiv arg \left [- \frac{V_{cd}V^*_{cb}}{V_{t
d}V^*_{tb}} \right ], \quad  \gamma \equiv arg\left [ - \frac{V_{ud}V^*_{ub}}{V_{cd}V^*_{cb}} \right ],  \\
\label{equ=angles}
\end{equation}

In the $B$ meson  sector there are many independent ways to measure \UT\  sides and angles, over-constraining this triangle.  To test the \SM  picture of \CP\ violation, we have to check  that the \UT\   is a closed triangle.  If experimental measurements  of  magnitudes of  \UT\ sides and  angles are inconsistent with a closed triangle, we have hints that New Physics (\NP) beyond the \SM contributes to \CP\ violation.   
 
The  \babar~\cite{babar} and Belle~\cite{belle} experiments, operating at the PEP-II and KEKB $B$-factories respectively, have provided in the last ten years very precise measurements  in the $B$ meson sector. The primary goal of these experiments was  the verification of the \SM\ description  of  \CP\ violation and this goal has been fully reached.  The observation of mixing-induced \CP\ violation in \psiKs\ decays~\cite{MixInd}, as well as in the charmless penguin-diagram dominated $\Bz \ra \etapr K^0$ decays~\cite{EtapKsBa, EtapKsBe}, and of direct \CP\ violation both in $ B^0\ra \pi^+ \pi^-$ and in  $B^0\ra K^+\pi^-$ decays \cite{dir}, are all in  agreement with \SM\ predictions.   

The \UT\  angle $\beta$ has been measured with high precision  from time-dependent  \CP\  asymmetries in neutral $B$ meson decays to \CP  eigenstates containing a charmonium and a $K^{(*)0}$ ~\cite{EtapKsBe, BestBetaBa} and will not be covered in this presentation.  

In the following, we present  some recent \babar\  measurements of the magnitudes of  \CKM\ elements $V_{ub}$ and $V_{cb}$,   and of the \UT angles $\alpha$ and $\gamma$.  

\section{Unitary Triangle  Sides}
 The measurement  of $|V_{ub}|$ and $|V_{cb}|$ has a  crucial  role in the test of the \SM.  In fact,  as shown in Fig.~1 and in Eq.~\ref{equ=sides}, the ratio $\frac{|V_{ub}|}{|V_{cb}|}$  is proportional to the length of \UT\ side that is opposite to the precisely measured angle $\beta$.  The value of this ratio constraints the upper  vertex of the \UT.  $|V_{ub}|$ and $|V_{cb}|$ appear  in the differential decay rate of semileptonic B decays  to charmless and charm final states, respectively.  Differently from hadronic $B$ decays, in such semileptonic $B$ decays hadronic and leptonic currents of the amplitude factorize. 

 Both  $|V_{ub}|$ and $|V_{cd}|$ can be  measured with an exclusive approach where the final state hadron is exclusively reconstructed and with an inclusive approach where all hadronic final states are summed.  In these two approaches  the hadronic current, difficult to evaluate, relies on different QCD calculations.  
  
\subsection{Inclusive Measurement  of $|V_{ub}|$ }
The magnitude of  $V_{ub}$ can be determined from inclusive  semileptonic $\Bbar$ decays  to charmless final states $X_u l \Nubar$, where $ l = e $ or $\mu$,  and $X_u$ is a hadronic system (without charm).   The real difficulty in this inclusive measurement comes from the 
overwhelming  charm background from $ \Bbar \ra X_c l \Nubar$ which has a rate fifty times larger and an event  topology very similar to signal.  

In a recent analysis~\cite{Sigamani} \babar, using the full dataset  of $467 \times 10^{6}$  \BBbar pairs,  has  measured  Partial Branching Fractions (PBF),  restricting  the analysis in selected regions  of the phase space where most  effective is the suppression of the charm background.  
The event selection uses a hadronic tag: in the sample of $\Upsilon(4S) \ra B \bar{B}$ one $B$  decaying into hadrons is 
fully reconstructed ($B_{tag}$) while the other $B$  ($B_{recoil}$) is identified by the precence of an electron or muon.
The $B_{tag}$ is reconstructed in many exclusive hadronic decays  $B_{tag} \rightarrow \bar{D}^{*} Y^{\pm}$, where the 
hadronic system $Y^{\pm}$ consists of hadrons and has a total charge of $\pm 1$. 
More than 1000 hadronic decay modes are reconstructed. 

In the $B_{recoil}$ rest-frame we require one lepton with momentum $p^*_l > 1 $ \gevc\  and the hadronic system  $X$ is reconstructed from charged particles and neutral clusters not associated to the $B_{tag}$ or the charged lepton. Neutrino is reconstructed from missing four-momentum in the whole event.  Requirements on several kinematic observables  were applied in different phase space regions to select the final signal events.  

PBFs are  measured in several regions of phase space and   are normalized to the total semileptonic branching fraction, thus reducing  several  systematic uncertainties. Considering the most inclusive measurement   (based only on the requirement 
$p^*  >1.0$  \gevc),  from a two-dimensional fit to the hadronic invariant mass  and to the leptonic invariant mass squared we measure :

\begin{equation}
\Delta {\cal B}(\Bbar  \ra X_u l \Nubar;  p^*_l > 1.0 \gevc) =  (1.80 \pm 0.13 \pm 0.15) \times 10^{-3} \,
\end{equation}

where the first uncertainty is statistical and the second  systematic.

To translate  the PBFs  measurements  into  $|V_{ub}|$ we need a theoretical extrapolation to the full space space, including 
perturbative and non perturbative QCD effects.  This extrapolation has been done using  four different models from Bosch, Lange, Neubert, and Paz (BLNP) \cite{Neubert},  Gambino, Giordano, Ossola, and Uraltsev (GGOU) \cite{Gambino}, Andersen and Gardi (DGE) \cite{Gardi}, and Aglietti, Di Lodovico, Ferrera, and Ricciardi (ADFR) \cite{Aglietti}. Making an arithmetic mean average of these four calculations, we obtain the result:

\begin{equation}
|V_{ub}| = (4.31 \pm 0.35) \times 10^{-3}
\end{equation}

This result with a total uncertainty  of about 8\% is comparable with Belle result \cite{InclBelle}.

\subsection{Exclusive Measurement of $ |V_{ub}| $ from $B \rightarrow (\pi, \rho) l \nu$ Decays}
\babar\   has measured $|V_{ub}| $ also with an exclusive approach in the charmless  semileptonic  decays $B\ra \pi l \nu$ and $B\ra \rho l \nu$ \cite{Vera}.  This analysis is based on a data sample of $377 \times 10^{6}$ \BBbar pair. In this exclusive analysis compared to the corresponding inclusive one
 we have a better control of the background but lower signal yields.  The  differential decay rate to the final state containing  the pseudoscalar meson 
 $\pi$ can be written in the form:
 
 \begin{equation}
 \frac{d\Gamma (B^0\ra \pi^- l^+  \nu)}{dq^2 d \cos\theta_{Wl}} = |V_{ub}|^2  \frac{G^2_F  p^3_{\pi}}{32 \pi^3} \sin^2 \theta_{Wl} | f_+(q^2)|^2 \, ,
 \label{eq:Rate}
 \end{equation}

where $G_F$ is the Fermi coupling constant,  $p_{\pi}$ is the momentum of the pion in the rest frame of the $B$ meson, $q^2$ is the momentum transfer squared from the $B$ meson to the 
final-state hadron (mass squared of the virtual $W$), 
$\theta_{Wl}$ is the angle of the charged-lepton momentum in the $W$ rest frame with respect to the direction of the $W$ boost from the $B$ rest frame,
and $f_+(q^2)$  is the form factor parameterizing the hadronic matrix element.  

 For the decays with the vector-meson $\rho$ in the final state the hadron 
matrix element is parameterized in terms of three form factors \cite{Vera}.

The four charmless semileptonic  decays $\Bz\ra \pi^- l^+ \nu$,  $B^+\ra \pi^0 l^+ \nu$, $B^0\ra \rho^- l^+ \nu$, and $B^+\ra \rho^0 l^+ \nu$ are reconstructed, requiring a high-momentum lepton  ($l = e$, $\mu$), a hadron ($\pi$, $\rho$), and a neutrino. The neutrino is reconstructed 
from the missing energy and momentum in the event.  All tracks and neutral clusters not associated to the signal must be consistent with a $B$ decay.
There are three types of background: continuum, \BBbar\ and other $B \ra X_u l \nu$ decay modes. $\BBbar$ is the largest source of background, in particular charmed semileptonic $B\ra X_c l \nu$.   Furthermore the isolation of the individual exclusive charmless decays from all the other $B \rightarrow X_u l \nu$ is difficult (they represent only $10\%$ of the total). The three types of background are suppressed using a neural network based on seven discriminating variables \cite{Vera}.

Branching fractions are extracted from extended binned maximum likelihood fit to \mes, \DE, and $q^2$. \mes\  is the 
beam-energy substituted $B$ mass and \DE\  is the difference between the reconstructed and expected energy of the $B$ candidate.
The four channels ($\pi^-$, $\pi^0$, $\rho^- $, and $\rho^0\ $) are fitted simultaneously  imposing isospin constraint.  Branching fraction results from 
this fit are:

\begin{eqnarray}
 {\cal B} (\Bz \ra \pi^- l^+\nu)  &  =  & (1.41 \pm 0.05 \pm 0.07) \times 10^{-4} \nonumber \\ 
{\cal B} (\Bz \ra \rho^- l^+\nu) &= & (1.75 \pm 0.15 \pm 0.27) \times 10^{-4} \, ,
\end{eqnarray} 

where the first uncertainty is statistical and the second systematic.

To extract $|V_{ub}|$ using Eq.~\ref{eq:Rate}  we need theoretical input for the form factor. For the $B\ra \pi l \nu$ partial  differential decay 
rates  $\Delta{\cal B}$ are measured in six bins of $q^2$ and results are compared (see Fig.~2) with calculations of  quark-model (ISGW2)~\cite{ISGW2},
QCD light-cone sum rules (LCSR1)~\cite{LCSR1},  (LCSR2)~\cite{LCSR2}, and unquanched lattice QCD calculations (HPQCD)~\cite{LQCD}.
The shape of the form factor is obtained  directly from the data. 

\begin{figure}[!h]
\begin{center}
\includegraphics[angle=0, scale=0.280] {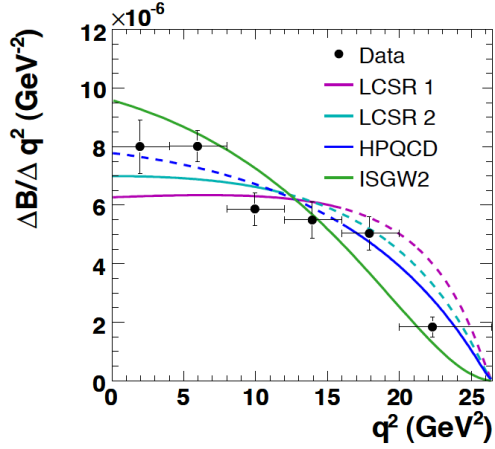} 
\caption{ Shape comparisons of measured partial branching fractions of $\Bz \ra \pi^- l^+ \nu$ to various form factor  
theoretical predictions which have been normalized to the measured total branching fraction. The dashed line represents the extrapolations of QCD predictions to the full $q^2$ range. }
\end{center}
\label{fig:PBFfit}
\end{figure}

The magnitude of $V_{ub}$ is extracted  with  two methods. In the first one $|V_{ub}|$ is obtained  by integration of form factor prediction  over the relevant $q^2$ interval using the relations:

$$
|V_{ub}| = \sqrt{\frac{\Delta {\cal B}(q^2_{min}, q^2_{qmax}) }{\tau_0 \Delta \zeta (q^2_{min}, q^2_{max})} }\, , \quad \Delta \zeta (q^2_{min}, q^2_{max}) = \frac{G^2_F}{24 \pi^3} \int_{q^2_{min} }^{q^2_{max}} p^3_{\pi} \left |  f_+(q^2) \right |^2 dq^2 \, ,
$$
where $\tau_0 = 1.530 \pm 0.009$ is the $B^0$ lifetime~\cite{Amsler}.

In Table~1 we show (first three rows) the extracted values of $|V_{ub}|$.  First quoted uncertainty  is  experimental and the second theoretical from the form-factor integral $\Delta \zeta$.  

\begin{table}[htb]
\caption{$ |V_{ub}|$ extracted from $B\ra \pi l \nu$ in various $q^2$ intervals and form factor calculations.  
In the last row $ |V_{ub}|$  measured in the simultaneous fit of \babar\  data to recent lattice calculations.  }
\label{tab:Vub}
\begin{center}
\begin{tabular}{lcccc}
\hline\hline
                           & $q^2$ Range    &  $\Delta {\cal B}$      & $\Delta \zeta$                          & $|V_{ub}| $\\ 
                           & $(GeV^2)$         & $(10^{-4})$                & $(ps^{-1})$                             &$(10^{-3})$ \\
\hline
LCSR 1            & $0 - 16$               & $1.10 \pm 0.07$       & $5.44 \pm 1.43$                   & $3.63 \pm 0.12 ^{+0.59}_{-0.40}$\\ 
LCSR 2            & $0 - 12$               & $0.88 \pm 0.06$       &  $4.00 ^{+1.01}_{-0.95}$     &  $3.78 \pm 0.13^{+0.55}_{-0.40}$\\
HPQCD            & $16 - 26.4$         & $0.32 \pm 0.03$       &  $2.02 \pm 0.55$                   & $3.21 \pm 0.17^{+0.55}_{-0.36}$\\
\hline
 FNAL/MILC    & $0 - 26.4$            & $1.41 \pm 0.09$       & $-$                                           & $2.95 \pm 0.31$\\
\hline\hline
\end{tabular}
\end{center}
\end{table}

\newpage
In the second method we do a simultaneous fit to the most recent lattice calculations and \babar\ data using the linear or quadratic BGL parameterization for  the full $q^2$ range~\cite{BGL}.   
In Fig.~3 we show results of such a fit. The solid line represents the quadratic (3 parameters + 1 normalization) BGL fit while the shaded 
region shows the uncertainty of the fitted function. The value of $|V_{ub}|$ extracted in this method using the normalization predicted by FNAL/MILC Collaboration~\cite{MILC} is is also shown (last row) in Table~1.  The quoted total uncertainty of  $10\, \%$  is  dominated by the theory uncertainty of $8.5\,\% $.

If we compare the \babar\ exclusive  and inclusive $|V_{ub}|$ determinations, we  see a discrepancy at the level of about $2.7 \, \sigma\ $.   A similar discrepancy at the  level of about $2.3 \, \sigma$ is also present in Belle results.

\begin{figure}[!h]
\begin{center}
\includegraphics[angle=0, scale=0.280] {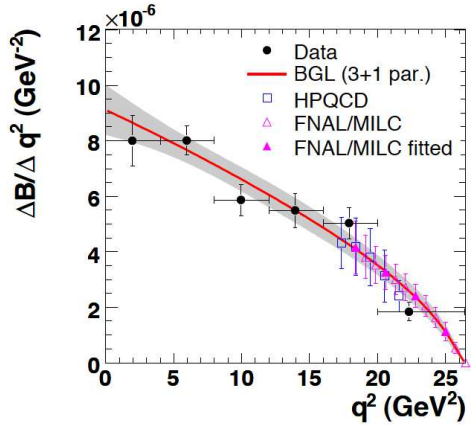} 
\caption{ Simultaneous fit  of quadratic BGL parameterization to data. The shaded band indicates  the uncertainty of the fitted function. }
\end{center}
\label{fig:bgl}
\end{figure}

\subsection{$|V_{ub}|$  and the Leptonic  $B^+ \rightarrow \tau^+ \nu_{\tau} $ Decay}
The purely leptonic  $B$ decay to $\tau \nu_{\tau}$ proceeds in \SM through $W$ boson annihilation with a branching fraction:

$$
{\cal B}(B^+ \rightarrow \tau^+ \nu_{\tau}) = f^2_B |V_{ub}|^2 \frac{G^2_F m_B m^2_{\tau}}{8 \pi} \left[1 - \frac{m^2_{\tau}}{m^2_B} \right]^2 \tau_{B^+} 
$$

The \SM estimate for this branching fraction is of the order of $10^{-4}$.  However contributions from  \NP\ scenarios ~\cite{Hou}  may enhance this expectation.

\babar\  has studied this decay mode both with a semileptonic tagging method~\cite{SemiLep} and with a  hadronic tag method~\cite{DeNardo}. We present  here results of the analysis based on hadronic tag. The  $B_{tag}$ candidates are reconstructed from $B^- \ra M^0 X^-$, where   $M^0 $ denotes a $D^{(*)0}$ 
or $J/\psi$, and $X^-$ is a hadronic system with total charge  -1.  The signal $ B$ candidate is reconstructed considering the most abundant decays  $\tau^+ \ra e^+ \nu \Nubar$, $\tau^+ \ra \mu^+ \nu \Nubar$, $\tau^+ \ra \pi^+ \nu$, and $\tau^+ \ra \rho^+ \nu$.  The most discriminating variable in this analysis is  $E_{extra}$, sum of the energies of neutral clusters not associated with the $ B_{tag}$ ( or with the $\pi^0$ from the $\tau^+ \ra \rho^+ \nu$) . Signal yield is extracted  from an extended  unbinned maximum likelihood  fit to all  four $\tau$ decay modes.  Table~2 summarizes fit results.

\begin{table}[htb]
\caption{Reconstruction efficiency $\epsilon$, branching fraction and significance (only statistical uncertainty)  from the fit to the four decay modes separately and constrained to the same branching fraction.}
\label{tab:TauNu }
\begin{center}
\begin{tabular}{lccc}
\hline\hline
Decay Mode                                  & $\epsilon \times 10^{-4}$       &   Branching Fraction $(\times 10^{-4})$ & Significance ($\sigma$) \\ [0.035in] \hline
$\tau^+ \ra e^+ \nu \Nubar$        & $2.73$                                       & $0.39^{+0.89}_{-0.79}$                             & $0.5$ \\ 
$  \tau^+ \ra \mu^+ \nu \Nubar $ & $2.92$                                      & $1.23^{+0.89}_{-0.80}$                              &  $1.6$\\
$\tau^+ \ra \pi^+ \nu$                   & $1.55$                                       & $4.0^{+1.5}_{-1.3}$                                     &  $3.3$\\
$ \tau^+ \ra \rho^+ \nu$               & $0.85$                                       & $4.3^{+2.2}_{-1.9}$                                     & $2.6$\\[0.035in]\hline
Combined                                      & $8.05$                                      & $1.80^{+0.57}_{-0.54}$                               &  $3.6$\\
\hline\hline
\end{tabular}
\end{center}
\end{table}

Including systematic uncertainties the branching fraction is ${\cal B}(B^+ \ra \tau^+ \nu_{\tau}) = (1.80^{+0.57}_{-0.54} \pm 0.26) \times 10^{-4} $. The null hypothesis (${\cal B}(B^+ \ra \tau^+ \nu_{\tau}) =0$) is excluded with a significance (including systematic uncertainty)  at the level of 3.3~$\sigma$.  

Combining this result and the other \babar\ measurement using a semileptonic tag and a statistical independent sample~\cite{SemiLep}, we obtain the result  ${\cal B}(B^+ \ra \tau^+ \nu_{\tau}) = (1.76  \pm 0.49) \times 10^{-4}$, where the uncertainty includes both statistical and systematic uncertainties.  

Results are consistent with the corresponding  Belle analyses using hadronic tag ~\cite{BelleHad} and semileptonic tag~\cite{BelleSemi}.

In a global fit excluding the branching fraction of $B^+ \ra \tau^+ \nu_{\tau}$, UTfit~\cite{UTfit}  finds for this branching fraction a value of 
$(0.79 \pm 0.07)
\times 10^{-4}$   while CKMfitter~\cite{CKMfitter} finds  $(0.786^{+0.179}_{-0.083}) \times 10^{-4}$. These  expectations (see Fig.~4) are about $2.5 \sigma$ lower that the experimental result.

\begin{figure}[!h]
\begin{center}
\includegraphics[angle=0, scale=0.30] {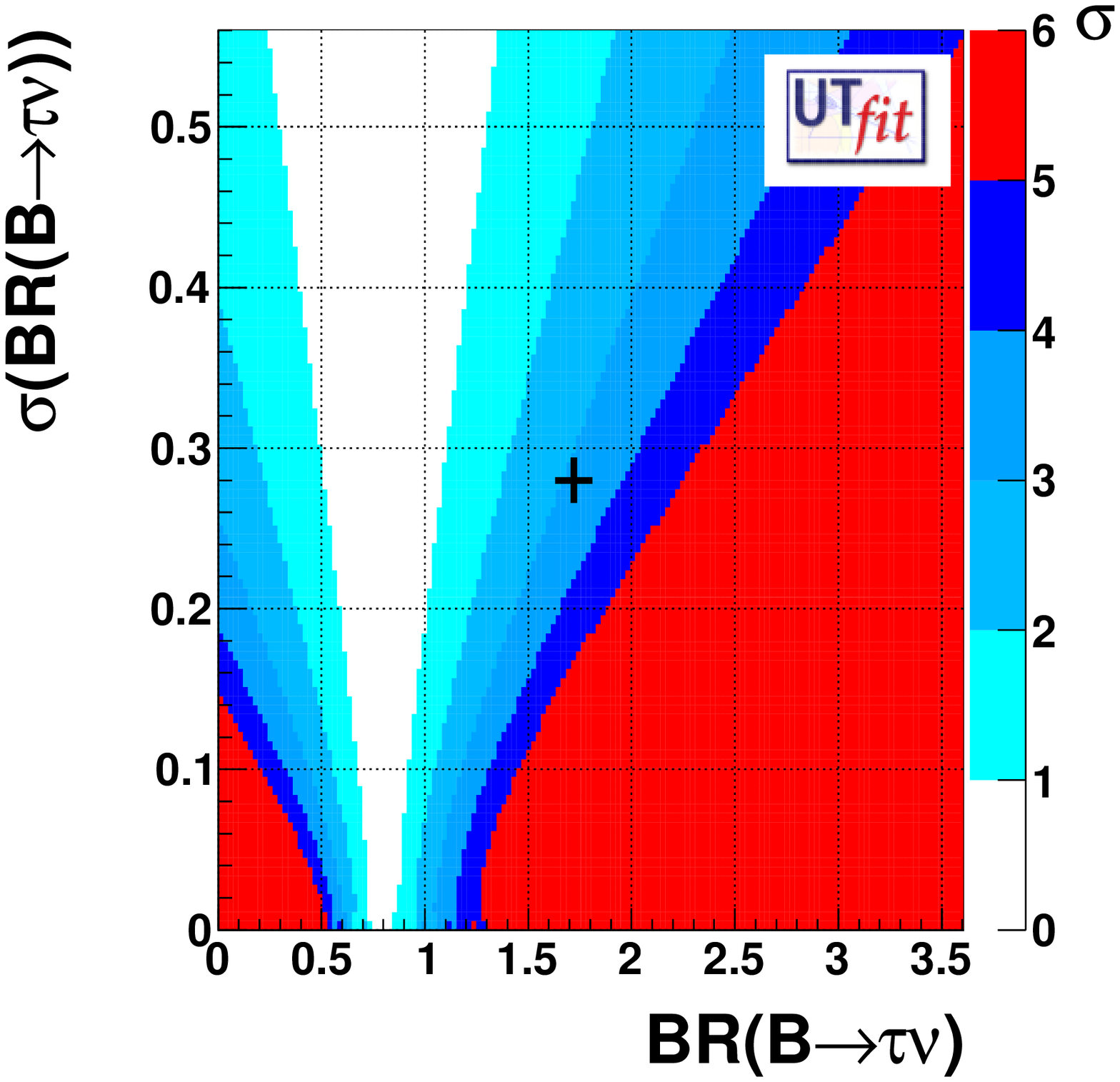} 
\includegraphics[angle=0, scale=0.41] {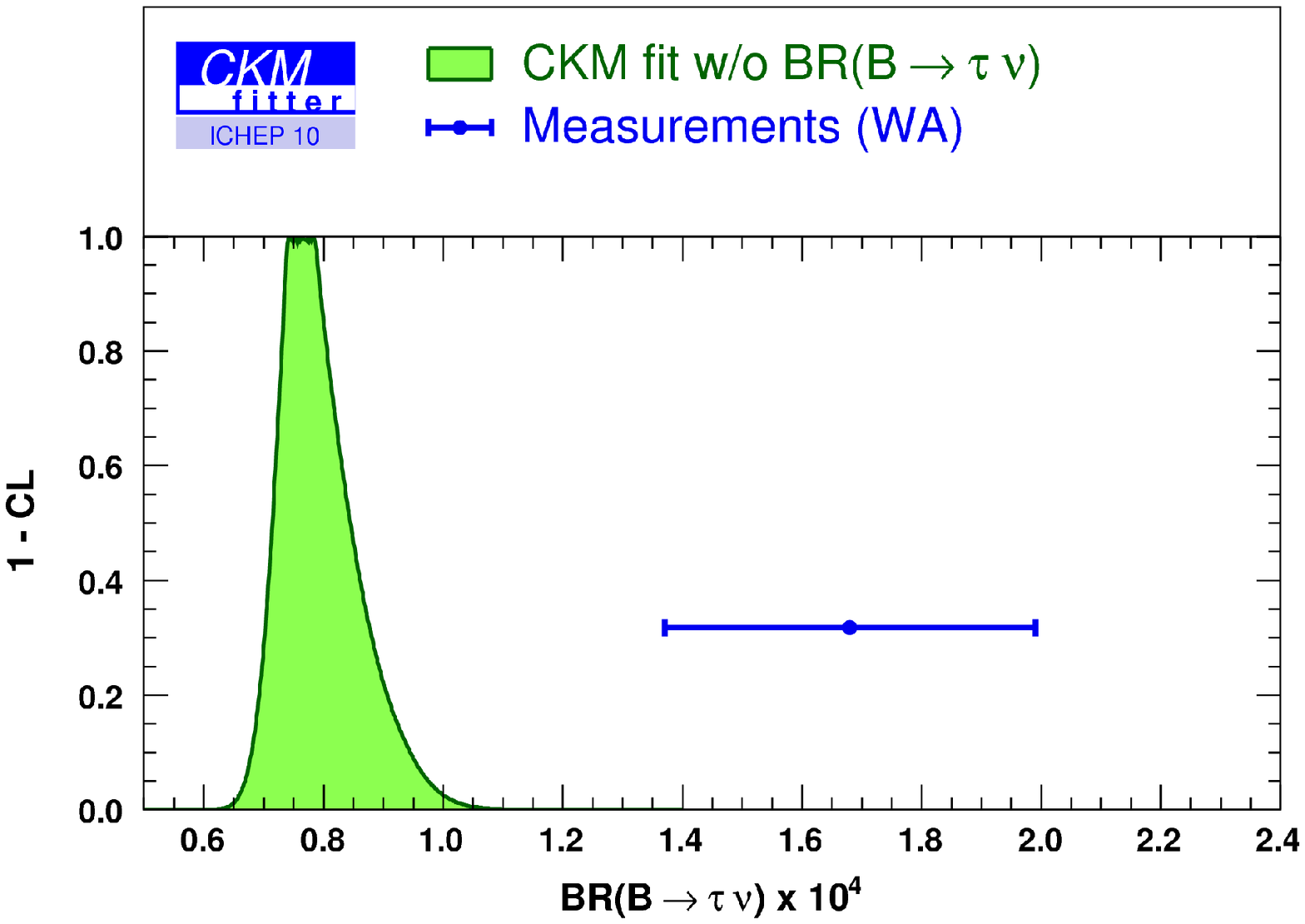} 
\caption{ Branching fraction of the decay $B^+ \ra \tau^+ \nu_{\tau}$ as predicted from a global fit by UTfit (on the left) and by CKMfitter (on the right)
compared to the experimental measurement.}
\end{center}
\label{fig:CKM}
\end{figure}

\subsection{Exclusive Measurement of $|V_{cb}| $ in $\Bbar \ra D l^- \Nubar_l$ Decays}
$|V_{cb}|$  has  been measured both in inclusive semileptonic  $B$ decays~\cite{semi}  and in exclusive decays $\Bbar \ra D l^- \Nubar_l$  and 
$\Bbar \ra D^* l^- \Nubar_l $~\cite{exclu} with l = e or $\mu$.  We present here a recent \babar\ measurement of  $|V_{cb}|$  from 
the differential decay rate of exclusive decays $\Bbar \ra D l^- \Nubar_l$~\cite{Vcb} :

\begin{equation} 
\frac{d\Gamma (\Bbar \ra D l \Nubar_l)}{d w} = \frac{G^2_F}{48 \pi^3 \hbar} m_D^3 (m_B + m_D)^2 (w^2 -1)^{\frac{3}{2}} |V_{cb}|^2 \mathcal{G}(w)
\end{equation} 

where $m_B$ and $m_D$  are the masses of $B$ and $D$ mesons, respectively. The variable $w$
 is the product of the $B$ and $D$ meson four-velocities,  $w = (m_B^2 + m_D^2 -  q^2)/(2 m_B m_D)$, where  $q^2  \equiv (p_B - p_D)^2$, and $p_B$ and $p_D$ are the four-momenta of the $B$ and $D$ mesons. $\mathcal{G}(w)$ is the form factor which is normalized to unity at zero recoil in heavy quark mass limit~\cite{wise}. 
 
 $|V_{cb}|$ is determined by extrapolating the differential decay rate to $w =1$. In this extrapolation the shape of the form factor is 
 needed. In this analysis  the parameterization proposed  in Ref.~\cite{caprini} has been adopted. Corrections to heavy quark limit have 
 been  calculated with unquenched~\cite{unquenched} and quenched~\cite{quenched}  lattice QCD. 
 
 Semileptonic signal $B$ events are searched for the recoil of fully reconstructed hadronic $B$ mesons ($B_{tag}$). They  are
  identified by their missing mass squared, calculated from the measured four-momenta of the particles in the event,  $m^2_{miss} = [ p_{\Upsilon(4S)} - p_{ B_{tag}}  -p_D -p_l  ]^2  $. This variable  peaks at zero for signal events. Signal yields are obtained from least-squares fit to the missing mass squared spectrum in ten equal-size intervals  of $w$ in the interval  $1<  w < 1.6$.  First fits   are done separately on the neutral and charged $\Bbar \ra D l^- \Nubar_l$ samples and then on the combined sample.  We show in Fig.~5 the measured $m^2_{miss}$ distributions and fit results (sum of the solid histograms) for two different $w$ intervals.
  
  \begin{figure}[!h]
\begin{center}
\includegraphics[angle=0, scale=0.28] {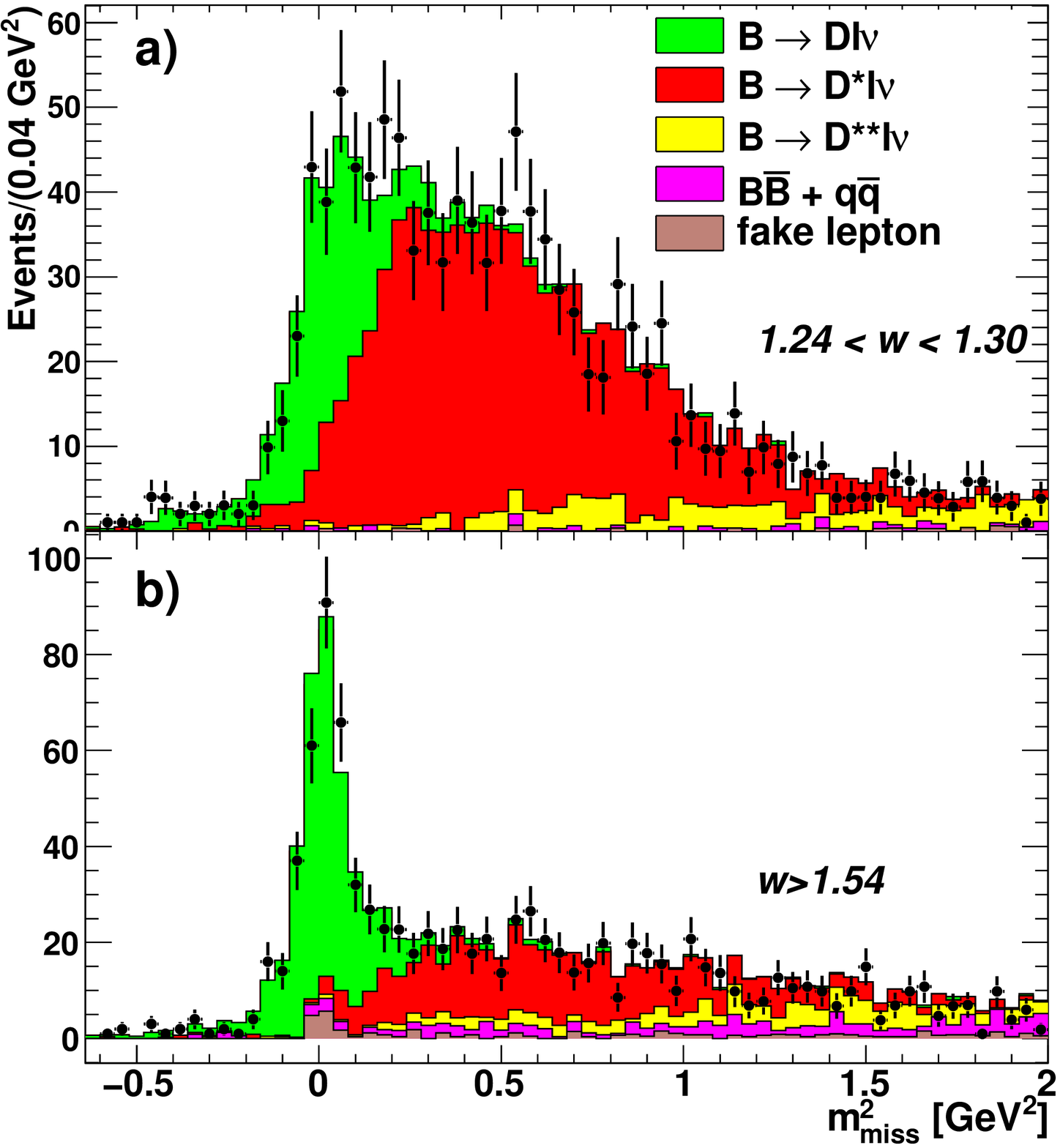}
\caption{ $m^2_{miss}$ distributions in two different $w$ intervals and fit results for $\Bm \ra \Dz l^- \Nubar$}.
\end{center}
\label{fig:Vcb1}
\end{figure}
  
We obtain    $ {\mathcal G}(1)  |V_{cb}|$ and  the form-factor slope $\rho^2$  from a least-squares fit to the $w$ distribution.
 We show in Fig.~6 data and fit results on the combined signal yields of   $\Bm \ra \Dz l^- \Nubar_l$ and $\Bzb \ra \Dp l^- \Nubar_l$.  The measured     $ {\mathcal G}(1)  |V_{cb}|$,  form-factor slope $\rho^2$ and branching fraction obtained from the fit to the combined $\Bzb/\Bm$ 
 sample are:

\begin{eqnarray}
 {\mathcal G}(1)  |V_{cb}|   &  =  & (42.3 \pm 1.9 \pm  1.4) \times 10^{-3} \nonumber \\
 \rho^2 & = & 1.20 \pm 0.09 \pm 0.04 \nonumber \\ 
{\cal B} (\Bbar \ra D  l^- \Nubar_l) &= & (2.15 \pm 0.06 \pm 0.09) \% 
\end{eqnarray} 

\begin{figure}[!h]
\begin{center}
 \includegraphics[angle=0, scale=0.25] {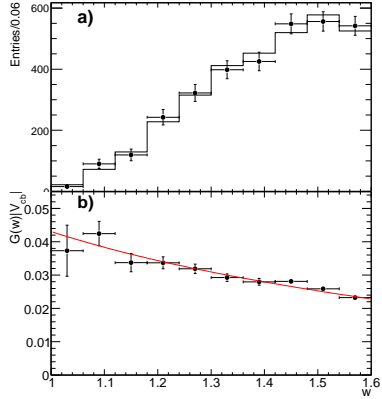}
\caption{ (a) w distribution of the signal yield of the combined sample.  Data points  are compared to the overall  fit results (histogram); (b)  $ {\mathcal G}(w)  |V_{cb}|$  distribution corrected for the reconstruction efficiency  with the fit superimposed. }
\end{center}
\label{fig:Vcb2}
\end{figure}

The extracted value of $|V_{cb}|$, applying an unquenched lattice calculation~\cite{unquenched}, is:

\begin{eqnarray}
 |V_{cb}| = (39.2 \pm 1.8 \pm 1.3 \pm 0.9) \times 10^{-3} \, ,
 \end{eqnarray}
  
where the uncertainties are statistical, systematic and theoretical (in  ${\mathcal G}(1)$ ),
respectively.

As for $|V_{ub}|$, $|V_{cb}|$ inclusive  results  tend to be higher than the exclusive results~\cite{HFAG}.

\section{Unitary  Triangle Angle $\alpha$}

$B^0$ decays to  to $\pi^+\pi^-$, $\rho^{\pm}\pi^{\mp}$, $\rho^+ \rho^-$,  and   $a_1(1260)^{\pm}\pi^{\mp}$  proceed dominantly 
through the $\bbar  \ra  \ubaru \dbar $ process and have  been  used  to measure the time-dependent
 \CP asymmetries and extract the angle $\alpha$.  In all these $B$  decay modes the presence of sizeable  loop (penguin) contributions introduces a distorsion (penguin pollution) in the measurement of $\alpha$. Instead of $\alpha $ one measures $\alpha_{eff}$. 
To take into account this distorsion several approaches have been proposed: isospin symmetry~\cite{ISO},  time-dependent Dalitz plot analysis~\cite{Quinn},  or approximate SU(3) flavor symmetry~\cite{Gross}.

We present here a recent  \babar\ update of the  measurement of  angle $\alpha$ in the  decay modes $B \ra \rho \rho$ and the first extraction of this angle  from the decay modes   \Nbtoappim.

\subsection {Angle $\alpha$ from $B\ra \rho \rho $ Decays  }
In the $B \ra \rho\ \rho\ $ decay modes the correction   
 $\Delta \alpha = \alpha - \alpha_{eff}$ has been obtained with an isospin analysis 
 involving  the $B \ra \rhoprhom$, \rhozrhoz, and \rhoprhoz\ decays.  In a recent   analysis~\cite{RhoRho} 
 \babar\  has updated with the full dataset  the measurement of the  branching fraction and longitudinal polarization 
 fraction $f_L$ in the $B \ra \rho^+  \rho^0$ decay, obtaining:
 
 \begin{eqnarray}
 {\cal B}(B^+ \ra \rho^+ \rho^0\ )   &  =  & (23.7  \pm 1.4 \pm  1.4) \times 10^{-6} \nonumber \\
f_L  &= & (0.950  \pm 0.015 \pm 0.006) \,
\end{eqnarray}

  These measured  values of  $ {\cal B}(B^+ \ra \rho^+ \rho^0\ ) $   and $f_L$ are higher than  those  of the previous \babar\ analysis~\cite{Prev} and this has an important effect in the isospin analysis.   ${\cal B}(B^+ \ra \rhoprhoz )$ represents in fact the common base of the two isospin triangles  of  the $B$ and the $\Bbar$  decays.  The large value of  ${\cal B}(B^+ \ra \rhoprhoz )$  flattens the two isospin triangles.   
   
An isospin analysis has been  performed using  the new   branching fraction and $f_L$   values  for the $B \ra \rho^+  \rho^0$ decay
 together with previous results  for $B^0 \ra \rho^+\rho^-$~\cite{RhopRhom}, and for $B^0 \ra \rho^0\rho^0$~\cite{RhozRhoz}.
  The four possible solutions of $\Delta \alpha$  are  now nearly degenerate while the eight-fold ambiguity on $\alpha$  degenerates  into a four-fold ambiguity with peaks  near $0^{\circ}$, $90^{\circ}$(two degenerate peaks), and $180^{\circ}$.  We take the solution for $\alpha$ near $90^{\circ}$ which is consistent with the global CKM  fits~\cite{UTfit, CKMfitter}.   Projections of the 1-CL scan  on $\alpha$ and $\Delta \alpha$  are shown in Fig.~7.

 \begin{figure}[htb]
\begin{center}
\includegraphics[scale=0.30] {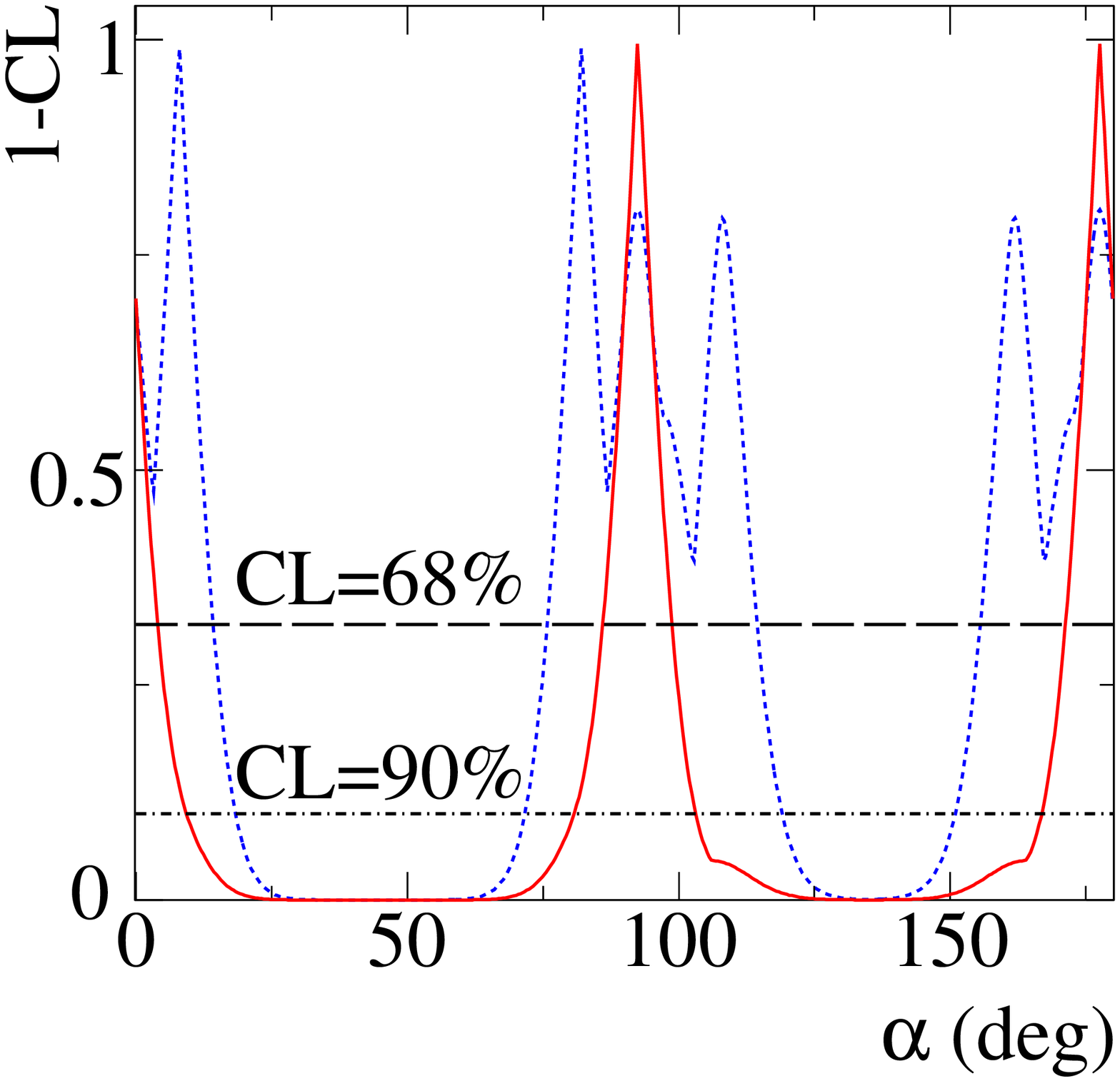} 
\hspace{1.5cm}
\includegraphics[scale=0.30] {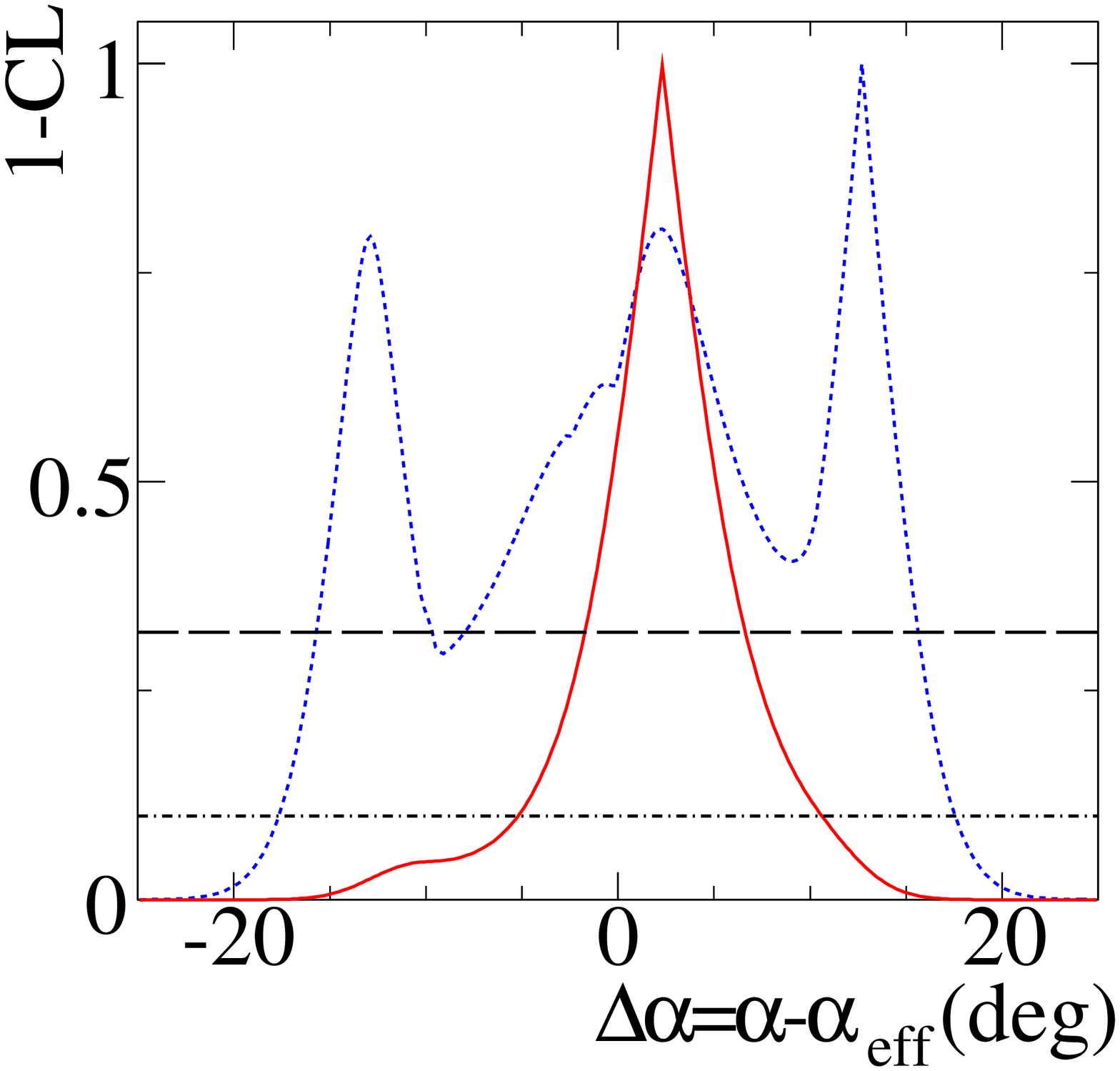} 
\vspace{0.5cm}
\caption{ Projections of 1-CL scan on  $\alpha$ and $\Delta \alpha$. The solid (dotted) curves show the results using the branching fraction ${\cal B}(B^+ \ra \rhoprhoz)$ measured in this analysis (prior to this analysis).  See supplemental material in Ref.~\cite{RhoRho}. }
\end{center}
\label{fig:RhoRho}
\end{figure}
  
  The solution for $\alpha$ and $\Delta \alpha$ at 68\% CL are $\alpha = (92.4^{+6.0}_{-6.5})^{\circ} $ and  $-1.8^{\circ} < \Delta \alpha < 6.7^{\circ} $. These results are significantly improved  compared  to those of the previous \babar\  analysis. This measurement is currently   the most precise single measurement of $\alpha$.

\subsection {Angle $\alpha$ from $B\ra a_1(1260)  \pi $ Decays }
The final states  in the $\Bz(\Bzb) \ra  a_1^{\pm}\pi^{\mp}$~\cite{Notation} decays  are  not  \CP eigenstates. So to 
extract $\alpha$ from  these  channels one has  to consider simultaneously \Bz(\Bzb) \ra \aunoppim and \Bz(\Bzb) \ra \aunompip~\cite{Aleksan}.  \babar\ has observed  these  decay modes~\cite{a1pi} and measured the time-dependent \CP asymmetries and $\alpha_{eff}$~\cite{CPa1pi}.  In these decay modes full isospin analysis or time-dependent Dalits plot approach to correct
for the distorsion $\Delta \alpha$ are not viable due to the limited statistics of available data samples, difficulties
because of the  four particles in the final states,  and uncertainties in the $a_1$ meson parameters and lineshape. 

Applying flavor SU(3) symmetry~\cite{Zupan} one can determine an upper  bound on  $\Delta \alpha =| \alpha -\alpha_{\rm eff}|$ by relating the \NNbtoappim decay 
rates  with  those of the $\Delta S =1$ transitions involving the same SU(3)  multiplet of \aunob, $B \ra \aunob K$ and $B \ra K_{1A} \pi$.  
Branching fractions of  $B \ra \aunob K$ have been already measured by \babar~\cite{a1K}.  The $K_{1A}$ meson is a nearly equal admixture 
of \Kuno\ and \Kunop\  resonances~\cite{Nakamura}. The rates of 
$B \ra K_{1A} \pi$ decays  can be derived from the decay rates of $B \rightarrow K_1(1270) \pi$ and  $B \rightarrow K_1(1400) \pi$.  

The $K_1(1270)$ and $K_1(1400)$ axial vector mesons  are broad resonances  with nearly equal masses. Both  mesons decay 
to the same final state $K\pi\pi$,  although through different intermediate states. However,  since the intermediate decays 
proceed almost at threshold, the available phase spaces  overlap and interference effects can be sizeable. The strategy of a recent  \babar\  analysis~\cite{BTOK1PI} 
relies on the reconstructed $K\pi\pi$ invariant mass  spectrum in the $\left[1.1,1.8\right]\,{\rm GeV}$ range to distinguish between $K_1(1270)$ and $K_1(1400)$, including interference effects in the signal model.  A two-resonance, six-channel $K$-matrix model~\cite{WA3}  in the P-vector approach~\cite{At} is used 
to describe the resonant $K\pi\pi$ system for the signal.

A MC technique is used to estimate a probability region  for the bound on $|\Delta \alpha|$ and the result is  $|\Delta \alpha|<11^{\circ}(13^{\circ})$ at 68\% (90\%) 
probability~\cite{BTOK1PI}.  Combining this bound on $\Delta \alpha$ and the measured $\alpha_{eff}$~\cite{CPa1pi}, we have  $\alpha = (79\pm 7\pm 11)^{\circ}$, where the first uncertainty is  statistical and systematic combined and the second uncertainty is  due to penguin pollution.

\section{Unitary Triangle Angle $\gamma$}
The angle $\gamma$ is the only \CP-violating parameters that  can be cleanly determined using solely tree-level $B$ decays.  In absence of penguin contribution, it  is almost largely unaffected by the presence of  \NP. We can access this angle
in the interference between the color-favored decay $B^- \ra D^{(*)0} K^{(*)-}$  ($b\ra c \ubar s$ transition)  and the color and CKM suppressed process $B^- \ra \Dbar^{(*)0} K^{(*)-}$ ($b\ra u\cbar s$ transition). Here $D$ refers to any admixture of  $D^0$ and its \CP-conjugate  $\Dbar^0$. The two interfering amplitudes differ by a factor $r_B e^{i ( {\delta}_B \pm \gamma)}$  where  $r_B$ is the magnitude of the ratio of the two amplitudes, and  $\delta_B $ is their relative strong phase.
 
Because of the limited available data sample and the small branching fractions of the target $B$ decay modes, angle $\gamma$
is the most difficult to measure and the less precisely known  UT angle.  

Several time-integrated  methods have been proposed to exploit this interference and extract angle  $\gamma$. The  most 
productive today are:  the Gronau, London, Wiler (GLW) method~\cite{GLW} where the Cabibbo-suppressed $D$ decays to \CP-eigenstates (such as $K^+ K^-$  or  $K^0_S \pi^0$); the Atwood, Dunietz, Soni (ADS) method~\cite{ADS} where $D$ is reconstructed in Cabibbo-favored and double Cabibbo-suppressed final states (such as $K^{\pm} \pi^{\mp}$); the  Giri, Grossman, Soffer, Zupan (GGSZ) method where the $D$ meson decays to thee-body self-conjugate final states (such as $K^0_S \pi^+ \pi^-$ or  $K^0_S K^+ K^-$) which are analyzed on a Dalitz plot.  \babar\ recently updated  $\gamma$ measurements with the GLW method~\cite{BGLW}, with the ADS method~\cite{BADS}, and  GGSZ method~\cite{BGGSZ}. We present here only  the  results obtained with the GGSZ method.

\subsection{Angle $\gamma$ Measured in a Dalitz Plot Analysis (GGSZ Method)}
In a recent  analysis based on the full \babar\ dataset ~\cite{BGGSZ} the angle $\gamma$ has been  measured following the GGSZ method. In this analysis the following  $B$ decay modes  are reconstructed:  $B^{\pm} \rightarrow D K^{\pm} $, $B^{\pm} \rightarrow D^* K^{\pm} $     
 ($D^* \rightarrow D \pi^0, D\gamma$) and $B^{\pm} \rightarrow D K^{* \pm}$ ($K^{*\pm} \rightarrow \K0S \pi^{\mp}$)  with   $D \ra K^0_S h^+ h^-$ ($h = \pi, K$).

The  three-body $D$ decays are studied on a Dalitz plot.  A simultaneous  extended unbinned maximum likelihood fit is done    
 to all the abovementioned $B$ decay modes, extracting signal and background yields
 together with the    \CP-violating parameters  $x_{\mp} \equiv  r_B \cos(\delta_B  \pm \gamma)$ and $y_{\mp} \equiv  r_B \sin(\delta_B  \pm \gamma)$ for $D K$, $D^* K$, and $D K^*$ final states. 
Using  all the measured observables 1-dimensional confidence intervals are  constructed following a frequentist approach.  

In Fig.~8 we show  1-CL as a function of $\gamma$ for the three B decay channels separately and their combination, including statistical and systematic uncertainties.  There is a single ambiguity in the weak and strong phases.  

 \begin{figure}[t]
\begin{center}
\includegraphics[angle=0, scale=0.3] {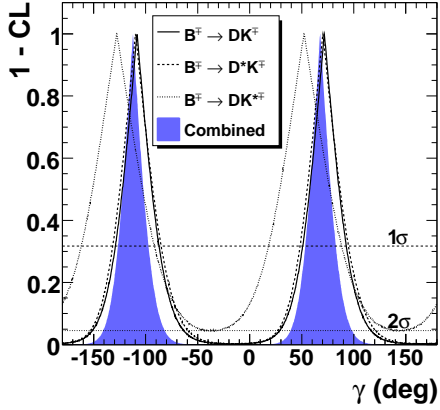} 
\caption{1-CL scan as a function of  $\gamma$  for $B^{\pm}  \ra DK^{\pm}$, $D^* K^{\pm}$, and $DK^{*\pm}$ decays separately, and combined. The dashed (upper) and dotter (lower) horizontal lines correspond to the one- and two-standard deviation intervals, respectively.  }
\end{center}
\label{fig:GAM-GGSZ}
\end{figure}

 The one- and two-standard deviation intervals
for $\gamma$  and for the three pairs of values  ($r_B$, $\delta_B$) ,  ($r^*_B$, $\delta^*_B$), and  ($\kappa r_s$, $\delta_s$) for the $DK$, $D^* K$ and $DK^*$ respectively,  are shown in Tab.~\ref{tab:TabGam}. The factor $\kappa = 0.9 \pm 0.1$ in the result for the decay $B^{\pm} \ra  DK^{*\pm}$ takes into account the $K^*$ finite width.

\begin{table}[htb]
\caption{The one- and two-standard deviation intervals for all relevant parameters. The first uncertainty is statistical. The second and third uncertainties inside \{ \} brackets are the symmetric uncertainty contributions to the total uncertainty  from the experimental and  neutral $D$ decay  amplitudes uncertainties.}
\label{tab:TabGam} 
\begin{center}
\begin{tabular}{lcc}
\hline\hline
Parameter            & $68.3\%$ CL  & $95.4\%$ CL \\ [0.035in] \hline
$\gamma$ $(^\circ)$        & $68^{+15}_{-14}$ $\{4,3\}$                  & $[39,98]$ \\ 
$r_B \ (\%)$           & $9.6 \pm 2.9$ $\{0.5,0.4\}$                 & $[3.7,15.5]$ \\
$r^*_B \ (\%)$         & $13.3^{+4.2}_{-3.9}$ $\{1.3,0.3\}$          & $[4.9,21.5]$ \\
$ \kappa r_s \ (\%)$         & $14.9^{+6.6}_{-6.2}$ $\{2.6,0.6\}$          & $< 28.0$ \\
\deltab   \ $(^\circ)$ & $119^{+19}_{-20}$ $\{3,3\}$                 & $[75,157]$ \\
\deltabst  \ $(^\circ)$ & $-82\pm21$ $\{5,3\}$                        & $[-124,-38]$ \\
\deltas  \  $(^\circ)$ & $111\pm32$ $\{11,3\}$                       & $[42,178]$ \\
\hline\hline
\end{tabular}
\end{center}
\end{table}

The extracted  central value  of $\gamma$  is $ (68 \pm 14 \pm 4 \pm 3)^{\circ}$  (modulo $180^{\circ}$).   It is inconsistent with no direct \CP violation ($\gamma = 0$) with a significance of 3.5 $\sigma$.
Results of this analysis are consistent with previous \babar\  results~\cite{BGGSZ-Pre} and with those of the  Belle Collaboration~\cite{GGSZ-Belle}.   

\newpage
\section{Summary and Conclusions}
We have presented results of some recent \babar\   measurements of the magnitudes of the CKM matrix elements  $V_{ub}$ and $V_{cb}$, and of the UT angles $\alpha$ and $\gamma$.  
In these   analyses \babar\ using final dataset  significantly decreased the uncertainties on the measurements of $|V_{ub}|$ and $|V_{cb}|$ , thanks to the increased dimension of the data sample,  improved experimental techniques and theoretical inputs.

A discrepancy at the level of about $2.7 \sigma$ is present between exclusive and inclusive $|V_{ub}|$ determinations. Similar discrepancy is also present in the results of the Belle analyses.  $|V_{cb}|$ inclusive results, as for $V_{ub}$, tend to be higher than the exclusive results. 
The inclusive and exclusive approaches however have different theoretical input: in the inclusive approach  parton level calculations need   perturbative corrections and non perturbative extrapolations while the exclusive approach is based on  the present lattice QCD and light cone sum rules understanding of form factors. It is not clear the source of these discrepancies which  may be due to some not well calibrated tool or be the  effect of unattributed uncertainties.

The updated measurement of the angle $\alpha$ in the $B\ra \rho \rho$ is significantly improved with respect to previous \babar\ and Belle results.  We have also presented a novel measurement of the angle $\alpha$ in  $B \ra a_1(1260) \pi$ decays. 

Finally we have reported on  a recent $\gamma$ measurement  in $B$ to $D^{(*)}K$ and $D K^{*\pm}$ decays using the  GGSZ method. This measurement, as well as all other $\gamma$ measurements at the $B$-factories,  is statistically limited. A precise $\gamma $ measurement 
is an important goal of next generation experiments on flavor physics.     

All the presented analyses, based on full or almost full \babar\ dataset, can be considered final.
Results of these analyses  as well as all \babar\  results are essentially in agreement with the expectations  of the \SM.
Both \babar\ and Belle experiments have found no clear effect which can be attributed to \NP.    The limited disagreement with \SM found in a few cases  may be explained either with improved calculations within the \SM or with \NP contributions. We need much more precise measurements to improve our sensitivity to \SM deviations and to \NP effects.
We expect a significant impact on flavor physics from  LHCb experiment~\cite{LHCb} and from the super flavor factories (superKEKB~\cite{SKB} and SuperB~\cite{superB}). 


\begin{thebibliography}{00}
\bibitem{Notazione}
In the literature another notation is also used, that is $\phi_1$, $\phi_2$, and $\phi_3$
for $\beta$, $\alpha$, and $\gamma$, respectively.
\bibitem{Notation}
For the $a_1(1260)$ meson we use the short notation $a_1$.
\end{thebibliography}
\end{document}